\documentclass[preprint,11pt,3p,A4paper,authoryear]{elsarticle}

\usepackage[whole]{bxcjkjatype}
\usepackage{newtxtext}
\usepackage{amsmath,amssymb,bm,bbm}
\usepackage{mathtools}
\usepackage{booktabs}
\usepackage{lineno}
\usepackage{color}
\usepackage{subcaption}
\usepackage{url}

\usepackage[ruled,lined,linesnumbered]{algorithm2e}
\DontPrintSemicolon

\usepackage[final]{showlabels}

\journal{Nuclear Engineering and Design}

\begin{document}

% \linenumbers

\begin{frontmatter}

    \title{Finite element model updating of building structures under seismic excitation:\\ A parallelized latent space--based Bayesian framework}
    
    \author[label1]{Taro Yaoyama\corref{cor1}}\ead{yaoyama@g.ecc.u-tokyo.ac.jp}
    \author[label1]{Sangwon Lee}
    \author[label2]{Minoru Matsubara}
    \author[label2]{Kenzo Kodera}
    \author[label2]{Takeshi Ugata}
    \author[label1]{Tatsuya Itoi}
    
    \cortext[cor1]{corresponding author}
    
    \affiliation[label1]{
        organization={Graduate School of Engineering, The University of Tokyo},
        addressline={7-3-1, Hongo}, 
        city={Bunkyo-Ku},
        postcode={113-8656}, 
        state={Tokyo},
        country={Japan}
    }
    
    \affiliation[label2]{
        organization={Taisei Corporation},
        addressline={1-25-1, Nishi-Shinjuku}, 
        city={Shinjuku-Ku},
        postcode={163-0606}, 
        state={Tokyo},
        country={Japan}
    }
    
    \begin{abstract}
        Enhancing seismic fragility and risk assessment of nuclear power plants relies on accurate prediction of reactor building responses to seismic hazards, which can be further improved through dynamic analysis of high-fidelity finite element (FE) models.
        However, FE models often exhibit non-negligible discrepancies from actual structures due to various sources of uncertainty, necessitating FE model updating with rigorous quantification of associated uncertainties.
        This paper presents a GPU-accelerated latent space--based Bayesian framework for FE model updating of building structures.
        In the proposed framework, high-dimensional structural response data (e.g., time histories or frequency response functions) are projected into a low-dimensional latent space using a multimodal variational autoencoder (MVAE), thereby enabling efficient and tractable likelihood evaluation without explicit modeling in the original observation space.
        Once trained, the surrogate enables amortized inference, allowing posterior sampling to be performed without additional simulator evaluations.
        We specifically employ a sequential Monte Carlo (SMC) sampler, whose population-based formulation allows parallel evaluation of the approximate likelihood on GPUs, resulting in computational efficiency and robustness against multimodal and complex posterior distributions.
        The proposed framework is validated through both numerical benchmarking and experimental data from a shaking table test of a reinforced concrete building structure.
        The results demonstrate that the method accurately estimates structural parameters with well-quantified uncertainties, while achieving fast and efficient inference through GPU-based parallelization, and enabling robust inference even in the presence of sparse observations that induce multimodal and highly complex posterior distributions.
    \end{abstract}
    

\end{frontmatter}

\section{Introduction} \label{sec:intro}

Enhancing seismic fragility and risk assessment of nuclear power plants depends on detailed response predictions for reactor buildings exposed to hazard-induced disturbances.
Especially when addressing seismic hazards, dynamic analysis of high-fidelity finite element (FE) models plays a critical role \citep{nakamuraN2010,choiB2022}.
However, non-negligible discrepancies often exist between model predictions and the observed behavior of actual buildings, owing to various factors,
e.g., measurement noise, variability in material properties, uncertainties introduced during the construction process,
over-simplification in the modeling of joints and boundaries, material degradation, and hazard-induced structural damage \citep{beckJL1998,sinhaJ2003,yaoyamaT2024a}.
To address this, Bayesian FE model updating \citep{simoenE2015,huangY2019ase,kiranRP2025} provides a rigorous framework for updating model parameters with quantified uncertainties.
This allows for robust detection of structural damage and material degradation, enhanced response predictions for future seismic hazards, and risk-informed maintenance activities.
In the field of nuclear engineering, several studies adopted Bayesian methods to update FE models of containment vessels \citep{songMY2024} and reinforced concrete (RC) building structures \citep{mengT2025}.

When considering high-fidelity FE models exposed to dynamic loading, the following challenges should be addressed.
First, model responses are typically observed in the form of time- or frequency-domain series that represent high-dimensional data sequences with highly correlated dimensions.
Formulating likelihood functions for such sequential data remains a significant challenge \citep{leeS2024asce,leeS2025mssp}.
Additionally, evaluating these likelihood functions involves calls to an expensive simulator (e.g., an ordinary differential equation solver),
making posterior inference computationally demanding.

To address these challenges, machine learning-aided approximate Bayesian approaches have gained increasing attention \citep{zengJ2025cmame,zengJ2025mssp,wangT2025,wangT2026}.
In such approaches, neural networks are trained on parameter–response pairs generated by a simulator and then used to approximate either the likelihood function or the posterior distribution.
This framework avoids explicit likelihood evaluations and substantially reduces the number of simulator calls during inference.
Recent studies have explored applications based on normalizing flows \citep{zengJ2025cmame,zengJ2025mssp,wangT2025} and diffusion models \citep{wangT2026}.
Similarly, our research group has developed a latent space-based Bayesian inference (LSBI) framework \citep{leeS2024asce,itoiT2024cmame,leeS2025mssp}.
In LSBI, an approximate likelihood is formulated in a low-dimensional latent space, which circumvents direct likelihood modeling in the original high-dimensional observation space.
When a multimodal variational autoencoder (MVAE) \citep{suzukiM2022} is employed for dimensionality reduction, posterior inference can be performed without additional simulator evaluations once the surrogate is trained, resulting in an efficient \textit{amortized} inference scheme \citep{itoiT2024cmame}.

Building upon these previous studies, this paper focuses on the sampling scheme for posterior inference.
While existing LSBI approaches \citep{leeS2024asce,itoiT2024cmame} rely on MCMC-based methods, we instead propose a GPU-enhanced parallel LSBI framework based on sequential Monte Carlo (SMC).
SMC is a particle-based sampling method in which a population of particles is progressively transported toward the target distribution.
This property enables robust inference even in the presence of multimodal or complex posteriors, which frequently arise in structural engineering applications due to limited sensor availability \citep{katafygiotisLS1998}.
Furthermore, its population-based structure naturally permits parallel evaluation of the approximate likelihood, making it well suited for GPU-accelerated computation.
In the following, we formulate the SMC-based LSBI framework (hereafter referred to as LSBI-SMC) and present a benchmarking study to demonstrate its robustness and efficiency for complex, multimodal posterior distributions in comparison with MCMC sampling schemes. Furthermore, in light of the lack of experimental validation in previous studies (e.g., \cite{itoiT2024cmame}), we investigate the applicability of the proposed framework to shake table test data of RC wall–frame building structures, which are structural types commonly used for reactor buildings in nuclear power plants.

\section{Latent space--based likelihood approximation revisited}

\begin{figure}[!t]
    \centering
    \includegraphics[width=0.8\linewidth]{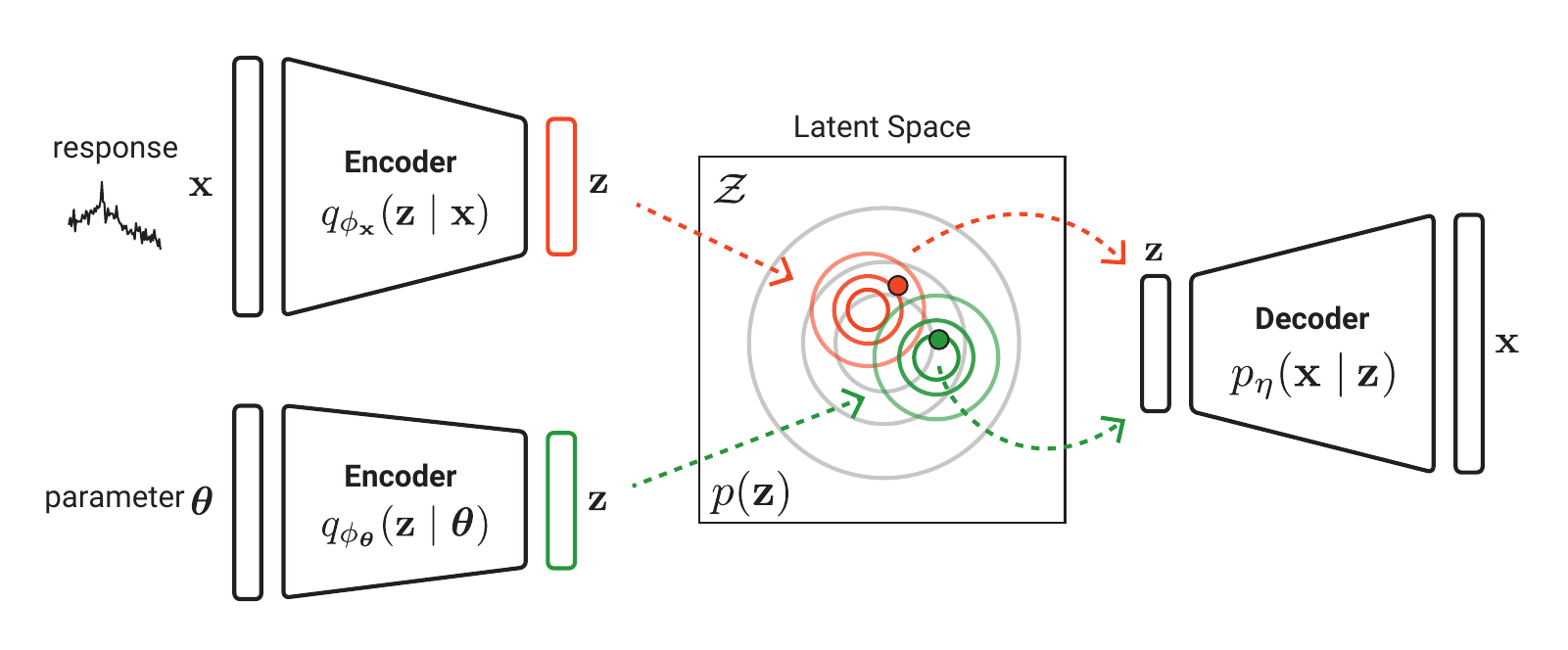}
    \caption{Schematic of Multimodal VAE architecture adopted in this study.}
    \label{fig:mvae}
\end{figure}

\subsection{Latent space--based Bayesian Inference with MVAE}

Consider an engineering system characterized by uncertain parameters $\bm{\theta} \in \mathbf{\Theta} \subseteq \mathbb{R}^{D}$, with responses of interest denoted by $\mathbf{x} \in \mathcal{X} \subseteq \mathbb{R}^{n_x}$.
We assume that the relationship between $\bm{\theta}$ and $\mathbf{x}$ is described by a simulator that induces a (possibly implicit) conditional distribution $p(\mathbf{x} \mid \bm{\theta})$.
In structural engineering applications, the simulator typically corresponds to a deterministic finite element (FE) analysis.
The data-generating process is thus modeled as
\begin{align}
    \mathbf{x} = h(\bm{\theta}) + \bm{\epsilon}, \quad \bm{\epsilon} \sim p(\bm{\epsilon}),
\end{align}
where $h(\cdot)$ denotes the FE simulator and $\bm{\epsilon}$ represents measurement noise.

For simplicity, we consider the case in which only a single observation $\mathbf{x}_\mathrm{obs} \in \mathcal{X}$ is available.
Extension to multiple observations is straightforward under the assumption of statistical independence.
Using Bayes' theorem, the posterior distribution of $\bm{\theta}$ conditioned on $\mathbf{x}_\mathrm{obs}$ is given by
\begin{align}
    p(\bm{\theta} \mid \mathbf{x}_\mathrm{obs})
    \propto L(\bm{\theta}; \mathbf{x}_\mathrm{obs}) \, p(\bm{\theta}),
\end{align}
where $L(\bm{\theta}; \mathbf{x}_\mathrm{obs}) := p(\mathbf{x}_\mathrm{obs} \mid \bm{\theta})$ denotes the likelihood function, and $p(\bm{\theta})$ and $p(\bm{\theta} \mid \mathbf{x}_\mathrm{obs})$ denote the prior and posterior probability density functions, respectively.
In engineering practice, such simulator-based inference often involves intractable or computationally prohibitive likelihood evaluations, owing to the high dimensionality of observed responses (e.g., seismic response histories) and the cost of repeated FE simulations.

To address this issue, we introduce a lower-dimensional latent variable $\mathbf{z} \in \mathcal{Z} \subset \mathbb{R}^{n_z}$ such that $\mathbf{x} \perp \bm{\theta} \mid \mathbf{z}$ and $n_z \ll n_x$, leading to the \textit{latent space--based likelihood approximation} \citep{leeS2024asce,itoiT2024cmame}:
\begin{align}
    L(\bm{\theta}; \mathbf{x}_\mathrm{obs}) 
    &= \int_\mathcal{Z} p(\mathbf{x}_\mathrm{obs} \mid \mathbf{z}) \, p(\mathbf{z} \mid \bm{\theta}) \, \mathrm{d} \mathbf{z} \nonumber \\
    &= c^{-1} \int_\mathcal{Z} \frac{p(\mathbf{z} \mid \mathbf{x}_\mathrm{obs}) \, p(\mathbf{z} \mid \bm{\theta})}{p(\mathbf{z})} \, \mathrm{d} \mathbf{z} \nonumber \\
    &\simeq c^{-1} 
    \int_\mathcal{Z} \frac{q_{\phi_\mathbf{x}} (\mathbf{z} \mid \mathbf{x}_\mathrm{obs}) \, q_{\phi_{\bm{\theta}}}(\mathbf{z} \mid \bm{\theta})}
    {p(\mathbf{z})} \,
    \mathrm{d}\mathbf{z}
    \label{eq:latent}
\end{align}
where $c = p(\mathbf{x}_\mathrm{obs}) = \int_\mathbf{\Theta} p(\mathbf{x_\mathrm{obs}} \mid \bm{\theta}) p(\bm{\theta}) \mathrm{d}\bm{\theta}$ denotes a normalizing constant;
$p(\mathbf{z})$ represents a prior distribution of $\mathbf{z}$ and is typically defined as a standard normal distribution, i.e., $p(\mathbf{z}) = \mathcal{N}(\mathbf{z} \mid \mathbf{0}, \mathbf{I}_{n_z})$, where $\mathbf{I}_n$ denotes an $n \times n$ identity matrix.
The densities $q_{\phi_\mathbf{x}}$ and $q_{\phi_{\bm{\theta}}}$ represent probabilistic encoders parameterized by $\phi_\mathbf{x}$ and $\phi_{\bm{\theta}}$, which provide shared embeddings from observed (or simulated) responses $\mathbf{x}$ and parameters $\bm{\theta}$.
This formulation avoids the need for explicit likelihood evaluation in the high-dimensional observation space.

Following \cite{itoiT2024cmame}, we specifically adopt probabilistic encoders from multimodal variational autoencoder (MVAE), leading to the MVAE--based LSBI framework.
MVAE is a neural network architecture designed to learn a shared latent representation across multiple data modalities (i.e., $\bm{\theta}$ and $\mathbf{x}$ in this study).
As illustrated in Figure~\ref{fig:mvae}, the architecture employed here consists of one decoder $p_\eta(\mathbf{x} \mid \mathbf{z})$ and two encoders $q_{\phi_\mathbf{x}}(\mathbf{z} \mid \mathbf{x})$ and $q_{\phi_{\bm{\theta}}}(\mathbf{z} \mid \bm{\theta})$.
The decoder $p_{\eta}$ represents a probabilistic generative process from a latent variable $\mathbf{z} \sim p(\mathbf{z}) := \mathcal{N}(\mathbf{z} \mid \mathbf{0}, \mathbf{I}_{n_z})$ to the response $\mathbf{x}$, and is modeled in a Gaussian form,
\begin{align}
    p_\eta(\mathbf{x} \mid \mathbf{z}) = \mathcal{N}(\mathbf{x} \mid \bm{\mu}_\eta(\mathbf{z}), \mathrm{diag}(\bm{\sigma}_\eta^2(\mathbf{z}))),
    \label{eq:peta}
\end{align}
where $\bm{\mu}_\eta(\cdot)$ and $\bm{\sigma}^2_\eta(\cdot)$ represent neural networks parameterized by $\eta$ and $\mathrm{diag}(\cdot)$ denotes a diagonal matrix constructed from a given vector.
The encoders $q_{\phi_\mathbf{x}}$ and $q_{\phi_{\bm{\theta}}}$ have the following forms,
\begin{align}
    q_{\phi_\mathbf{x}}(\mathbf{z} \mid \mathbf{x}) &=
    \mathcal{N}(\mathbf{z} \mid \bm{\mu}_\mathrm{\phi_\mathbf{x}}(\mathbf{x}), \mathrm{diag}(\bm{\sigma}^2_\mathrm{\phi_\mathbf{x}}(\mathbf{x}))), \label{eq:qphi1} \\
    q_{\phi_{\bm{\theta}}}(\mathbf{z} \mid \bm{\theta}) &=
    \mathcal{N}(\mathbf{z} \mid \bm{\mu}_\mathrm{\phi_{\bm{\theta}}}(\bm{\theta}), \mathrm{diag}(\bm{\sigma}^2_\mathrm{\phi_{\bm{\theta}}}(\bm{\theta}))).
    \label{eq:qphi2}
\end{align}
where $\bm{\mu}_{\phi_\mathbf{x}}, \bm{\sigma}^2_{\phi_\mathbf{x}}$ and $\bm{\mu}_{\phi_{\bm{\theta}}}, \bm{\sigma}^2_{\phi_{\bm{\theta}}}$ are neural networks parameterized by weights $\phi_\mathbf{x}$ and $\phi_{\bm{\theta}}$, respectively.
We accordingly define the MVAE-based likelihood approximation as
\begin{align}
    \widehat{L}_{\phi_\mathbf{x},\phi_{\bm{\theta}}}(\bm{\theta}; \mathbf{x}) :=
    \int_\mathcal{Z} \frac{q_{\phi_\mathbf{x}} (\mathbf{z} \mid \mathbf{x}_\mathrm{obs}) \, q_{\phi_{\bm{\theta}}}(\mathbf{z} \mid \bm{\theta})}
    {p(\mathbf{z})} \,
    \mathrm{d}\mathbf{z},
\end{align}
Notably, as we adopt Gaussian forms for the encoders $q_{\phi_\mathbf{x}}(\mathbf{z} \mid \mathbf{x})$, $q_{\phi_{\bm{\theta}}}(\mathbf{z} \mid \bm{\theta})$ and prior $p(\mathbf{z})$, the integral in Equation (\ref{eq:latent}) reduces to a closed form \citep{leeS2024asce} and can readily be evaluated.

\subsection{Training procedure for the MVAE-based likelihood approximation}

In conventional (M)VAEs, the loss function is typically formulated to reconstruct the input data itself through a latent space, i.e., $\mathbf{x} \rightarrow \mathbf{z} \rightarrow \mathbf{x}$.
In contrast, to explicitly account for stochasticity in the simulator $p(\mathbf{x} \mid \bm{\theta})$, we consider two independent samples $(\mathbf{x}, \tilde{\mathbf{x}})$ drawn from $p(\mathbf{x} \mid \bm{\theta})$ (or equivalently, a noise-free response $h(\bm{\theta})$ and a noisy response $h(\bm{\theta}) + \bm{\epsilon}$).
Accordingly, the model is trained to perform stochastic reconstructions of the form $\mathbf{x} \rightarrow \mathbf{z} \rightarrow \tilde{\mathbf{x}}$, as well as cross-modal mappings from parameters to responses, i.e., $\bm{\theta} \rightarrow \mathbf{z} \rightarrow \tilde{\mathbf{x}}$.

The resulting training procedure is as follows.
First, we draw parameter samples from a predefined proposal distribution $\pi(\bm{\theta})$ (typically set to the prior $p(\bm{\theta})$), i.e., $\bm{\theta}^{(n)} \sim \pi(\bm{\theta})$ for $n = 1, ..., N_\mathrm{train}$.
For each $\bm{\theta}^{(n)}$, we call a simulator $p(\mathbf{x} \mid \bm{\theta})$ to generate two independent samples $(\mathbf{x}^{(n)}, \tilde{\mathbf{x}}^{(n)})$.
In the case of a deterministic simulator with additive observation noise, this reduces to
\begin{align}
    \mathbf{x}^{(n)} &= h(\bm{\theta}^{(n)}), \\
    \tilde{\mathbf{x}}^{(n)} &= h(\bm{\theta}^{(n)}) + \bm{\epsilon}^{(n)}, 
    \quad \bm{\epsilon}^{(n)} \sim p(\bm{\epsilon}),
\end{align}
where the noise model $p(\bm{\epsilon})$ is specified based on prior knowledge or engineering expertise.
Using the resulting dataset $\mathcal{D} = \{(\bm{\theta}^{(n)}, \mathbf{x}^{(n)}, \tilde{\mathbf{x}}^{(n)})\}$, we train neural networks $(\eta, \phi_\mathbf{x}, \phi_{\bm{\theta}})$ by minimizing the following loss function:
\begin{align}
    \mathcal{L}(\eta, \phi_\mathbf{x}, \phi_{\bm{\theta}})
    = \mathcal{L}_\mathrm{recon}(\eta, \phi_\mathbf{x})
    + \mathcal{L}_\mathrm{pred}(\eta, \phi_{\bm{\theta}})
    + \mathcal{L}_\mathrm{regul}(\phi_\mathbf{x}, \phi_{\bm{\theta}}),
    \label{eq:loss}
\end{align}
where the reconstruction, prediction, and regularization terms are defined as
\begin{align}
    \mathcal{L}_\mathrm{recon}(\eta, \phi_\mathbf{x})
    &= -\mathbb{E}_{q_{\phi_\mathbf{x}}(\mathbf{z} \mid \mathbf{x})}
    \left[
    \log p_\eta(\tilde{\mathbf{x}} \mid \mathbf{z})
    \right], \\
    \mathcal{L}_\mathrm{pred}(\eta, \phi_{\bm{\theta}})
    &= -\mathbb{E}_{q_{\phi_{\bm{\theta}}}(\mathbf{z} \mid \bm{\theta})}
    \left[
    \log p_\eta(\tilde{\mathbf{x}} \mid \mathbf{z})
    \right], \\
    \mathcal{L}_\mathrm{regul}(\phi_\mathbf{x}, \phi_{\bm{\theta}})
    &= D_\mathrm{KL}\!\left(q_{\phi_\mathbf{x}}(\mathbf{z} \mid \mathbf{x}) \,\|\, p(\mathbf{z})\right)
    + D_\mathrm{KL}\!\left(q_{\phi_{\bm{\theta}}}(\mathbf{z} \mid \bm{\theta}) \,\|\, p(\mathbf{z})\right) \nonumber \\
    &\quad + \alpha \Big[
    D_\mathrm{KL}\!\left(q_{\phi_\mathbf{x}}(\mathbf{z} \mid \mathbf{x}) \,\|\, q_{\phi_{\bm{\theta}}}(\mathbf{z} \mid \bm{\theta})\right)
    + D_\mathrm{KL}\!\left(q_{\phi_{\bm{\theta}}}(\mathbf{z} \mid \bm{\theta}) \,\|\, q_{\phi_\mathbf{x}}(\mathbf{z} \mid \mathbf{x})\right)
    \Big].
\end{align}
Here, $\mathbb{E}_p[\cdot]$ denotes the expectation with respect to a probability distribution $p$, and $D_\mathrm{KL}(p \,\|\, q)$ denotes the Kullback--Leibler divergence between distributions $p$ and $q$.  
The hyperparameter $\alpha$ controls the strength of the consistency regularization between the two encoders, and is set to $\alpha = 5$ in this study.

\section{Parallelized LSBI framework with Sequential Monte Carlo sampler}

\subsection{Sequential Monte Carlo}

\begin{figure}[!t]
    \centering
    \includegraphics[width=1.0\linewidth]{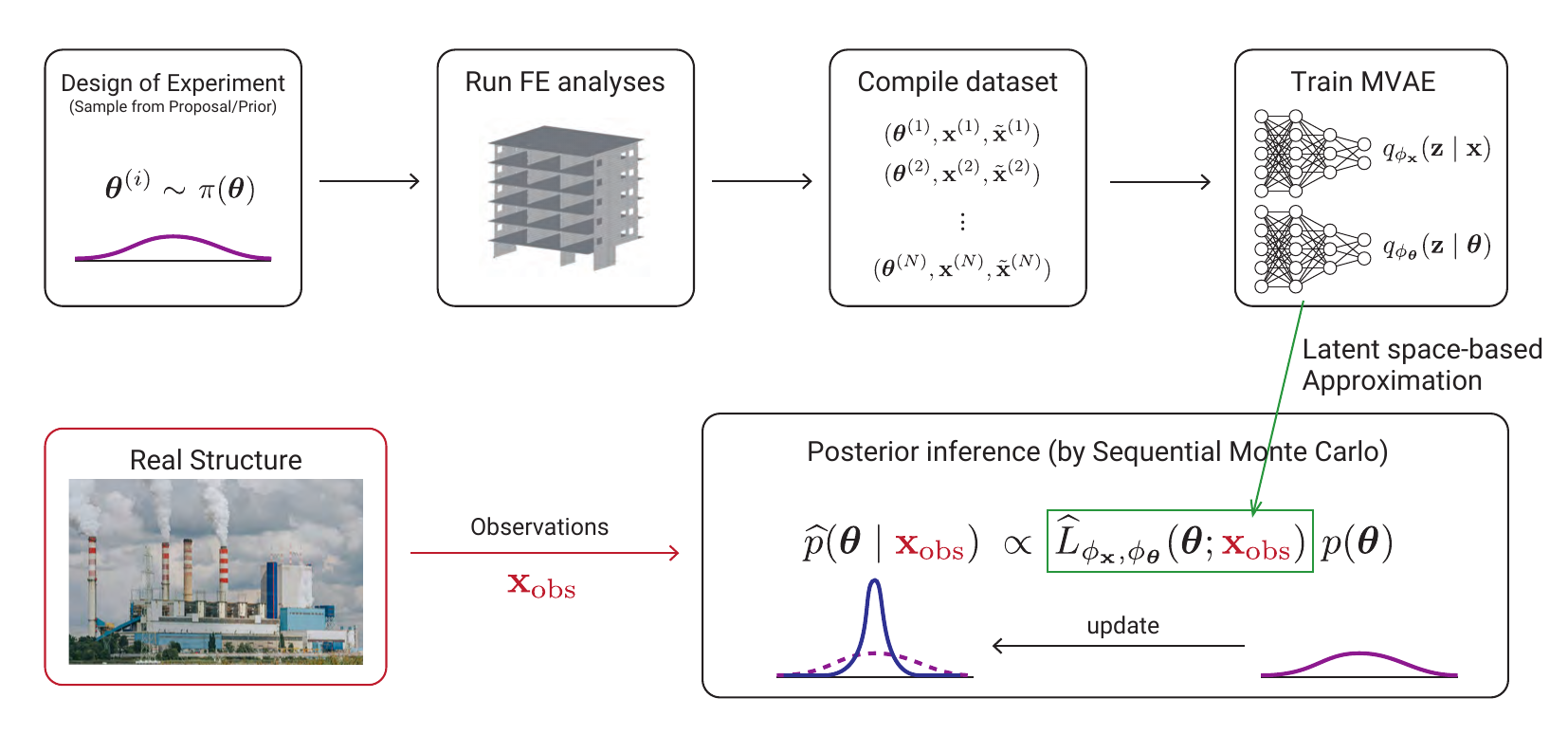}
    \caption{The overview of the proposed framework: LSBI-SMC.}
    \label{fig:flowchart}
\end{figure}

To infer the posterior $\widehat{p}(\bm{\theta} \mid \mathbf{x}_\mathrm{obs}) \propto \widehat{L}_{\phi_\mathbf{x},\phi_{\bm{\theta}}}(\bm{\theta}; \mathbf{x}_\mathrm{obs}) \, p(\bm{\theta})$, we adopt SMC sampler \citep{chingJ2007,betzW2016,carreraB2024}.
The general algorithm of SMC is formulated as follows.
We initially place a population of samples (particles) drawn from the prior $p(\bm{\theta})$, and gradually push them to the target posterior via the suite of \textit{intermediate} probability densities,
\begin{align}
    p_t(\bm{\theta}) = c^{-1} ~ \{\widehat{L}_{\phi_\mathbf{x},\phi_{\bm{\theta}}}(\bm{\theta};  \mathbf{x}_\mathrm{obs})\}^{\beta_t - \beta_{t-1}} p_{t-1}(\bm{\theta}), ~~~
    \text{for} ~ t = 1, ..., T.
\end{align}
$p_0(\bm{\theta}) = p(\bm{\theta})$ and the resulting distribution $p_T(\bm{\theta})$ corresponds to $\widehat{p}(\bm{\theta} \mid \mathbf{x}_\mathrm{obs})$.
$\beta_t$ is a tempering parameter that starts at $\beta_0 = 0$ and monotonically increases until $\beta_T = 1$.
At each level $t$, $\beta_t$ is determined based on the particles $\{\bm{\theta}_{t-1}^{(i)}\}_{i=1}^{N_\mathrm{s}}$ at the previous level by the following rule:
\begin{align}
    \beta_t = \mathrm{argmin}_{\beta \in (\beta_{t-1}, 1]}
    \left\{N_\mathrm{eff}\left(\beta; \{\bm{\theta}_{t-1}^{(i)}\}_{i=1}^{N_\mathrm{s}}\right) - \gamma N_\mathrm{s}\right\}
\end{align}
where $N_\mathrm{eff}$ denotes the effective sample size (ESS) described as
\begin{align}
    N_\mathrm{eff}\left(\beta; \{\bm{\theta}^{(i)}_{t-1}\}_{i=1}^{N_\mathrm{s}} \right) =
    \frac{\left( \sum_{i=1}^{N_\mathrm{s}} w_{t}^{(i)}(\beta) \right)^2}{\sum_{i=1}^{N_\mathrm{s}} {w_{t}^{(i)}(\beta)}^2}
\end{align}
with $w_{t}^{(i)}(\beta) = \widehat{L}_{\phi_\mathbf{x},\phi_{\bm{\theta}}}(\bm{\theta}_{t-1}^{(i)} \mid \mathbf{x}_\mathrm{obs})^{\beta - \beta_{t}}$ denoting the weight of each particle for a specified $\beta$.
The hyperparameter $\gamma$ controls the degree of transition with $\gamma N_\mathrm{s}$ representing the target ESS.

The update of particles at each level $t$ consists of two steps: ``resample'' and ``move''.
In the former step, $N_\mathrm{s}$ particles are resampled according to the weights $w_{t}^{(i)}(\beta_t)$.
In the latter step, MCMC updates with a Markov transition kernel $\mathcal{K}(\bm{\theta}, \cdot)$ are performed for each particle.
We specifically adopt the random walk Metropolis--Hastings (RWMH) kernel, in which a candidate position is sampled according to $\bm{\theta}^\ast \sim \mathcal{K}(\bm{\theta}, \cdot)$ and evaluated according to the acceptance probability given by
\begin{align}
    \eta = \min \left\{1, \, \frac{\mathcal{K}(\bm{\theta}^\ast, \bm{\theta})}{\mathcal{K}(\bm{\theta}, \bm{\theta}^\ast)}\frac{p_t(\bm{\theta}^\ast)}{p_t(\bm{\theta})} \right\}.
\end{align}
The transition kernel is typically considered as a symmetric distribution such as Gaussian, $\mathcal{K}(\bm{\theta}, \cdot) = \mathcal{N}(\cdot \mid \bm{\theta}, \mathbf{\Sigma})$, resulting in the acceptance probability with a reduced form, $\eta = \min\{1, ~ p_t(\bm{\theta}^\ast)/p(\bm{\theta})\}$.
\cite{chingJ2007} proposed the use of a Gaussian proposal with a covariance,
\begin{align}
    \mathbf{\Sigma}_t = b^2 \cdot \sum_{i=1}^{N_\mathrm{s}}
    \frac{w_t^{(i)}}{S_t \cdot N_\mathrm{s}} (\bm{\theta}_t^{(i)} - \overline{\bm{\theta}}_t) (\bm{\theta}_t^{(i)} - \overline{\bm{\theta}}_t)^\top
    \label{eq:ching_proposal}
\end{align}
where $S_t = \sum_{i=1}^{N_\mathrm{s}} \left. w_t^{(i)} \middle/ N_\mathrm{s} \right.$ and
$\overline{\bm{\theta}}_t = \left. \sum_i w_t^{(i)} \bm{\theta}_t^{(i)} \middle/ \sum_i w_t^{(i)} \right.$.
In this study, we set hyperparameters as $\gamma = 0.8$ and $b = 0.2$.

\subsection{Proposed framework}

The proposed framework, LSBI-SMC, is summarized as follows.
\begin{enumerate}
    \item Draw parameter samples from a predefined proposal distribution $\pi(\bm{\theta})$ (typically set to the prior distribution $p(\bm{\theta})$): $\bm{\theta}^{(n)} \sim \pi(\bm{\theta})$ for $n = 1, \ldots, N_\mathrm{train}$.
    \item For each $\bm{\theta}^{(n)}$, generate two independent response samples from the simulator: $\mathbf{x}^{(n)}, \tilde{\mathbf{x}}^{(n)} \sim p(\mathbf{x} \mid \bm{\theta}^{(n)})$, resulting in a training dataset $\mathcal{D} = \{(\bm{\theta}^{(n)}, \mathbf{x}^{(n)}, \tilde{\mathbf{x}}^{(n)})\}_{n=1}^{N_\mathrm{train}}$.
    \item Using $\mathcal{D}$, train the neural networks $(\eta, \phi_\mathbf{x}, \phi_{\bm{\theta}})$ by minimizing the loss function $\mathcal{L}$ in Equation~(\ref{eq:loss}) to construct the approximate likelihood function $\widehat{L}_{\phi_\mathbf{x}, \phi_{\bm{\theta}}}(\bm{\theta}; \mathbf{x})$.
    \item For a given observation $\mathbf{x}_\mathrm{obs}$, run the SMC sampler to draw posterior samples from $\widehat{p}(\bm{\theta} \mid \mathbf{x}_\mathrm{obs}) \propto \widehat{L}_{\phi_\mathbf{x}, \phi_{\bm{\theta}}}(\bm{\theta}; \mathbf{x}_\mathrm{obs}) \, p(\bm{\theta})$.
\end{enumerate}
This procedure is illustrated in Figure~\ref{fig:flowchart}.

In this framework, the surrogate likelihood $\widehat{L}_{\phi_\mathbf{x}, \phi_{\bm{\theta}}}(\bm{\theta}; \mathbf{x}_\mathrm{obs})$ enables amortized inference without additional simulator evaluations.
The population-based SMC sampler further allows efficient GPU-based parallel evaluation of the surrogate likelihood, resulting in improved computational and sampling efficiency.

\section{Benchmarking example}

\begin{figure}[!t]
    \centering
    \includegraphics[width=0.50\linewidth]{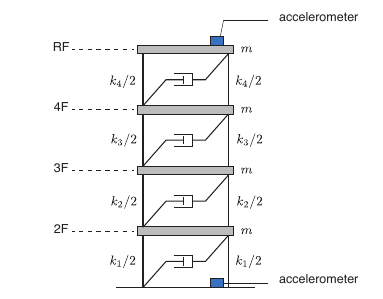}
    \caption{The target structure in the benchmarking example: a shear-type 4DOF building model.}
    \label{fig:shear4dof}
\end{figure}

\subsection{Target structure and data}

The target structure is a shear-type 4DOF building structure as illustrated in Figure~\ref{fig:shear4dof}.
We assume that the masses for all floors have an identical value of $m$ (kg).
Rayleigh damping is adopted with a damping ratio of 0.02 both at $1.0$ and $20.0$ Hz.
The stiffness values of all stories are parameterized as $k_i = \theta_i \, \overline{k}$ with $\overline{k} = 1000 \cdot m$ ($\text{s}^{-2}$) denoting a representative value.
The coefficients $\bm{\theta} := \{\theta_i\}_{i=1}^4$ are uncertain parameters to be updated with uniform priors $\theta_i \sim \mathcal{U}(\cdot \mid [0.33, 3.00]) ~ (i = 1, ..., 4)$, where $\mathcal{U}(\cdot \mid S)$ denotes the probability density function of a uniform distribution defined over the set $S$.
In this example, we assume that acceleration measurements are available only at the base and roof-top.
Under this assumption, the parameter $\bm{\theta}$ is not globally identifiable and permits several equivalent solutions \citep{katafygiotisLS1998}.
We set $\bm{\theta} = \{1, 1, 1, 1\}$ as ground truth values and apply the algorithm presented in \cite{katafygiotisLS1998} to identify the other three solutions that reproduce the identical roof-top response, as listed in Table~\ref{tab:sols}.
This implies that our target posterior exhibits severe multimodality.

\begin{table}[b]
    \centering
    \caption{Equivalent parameters that reproduce the identical roof-top response in the 4DOF example.}\label{tab:sols}
    \fontsize{8truept}{10truept}\selectfont
    \begin{tabular*}{\textwidth}{@{\extracolsep{\fill}}rrrrr}
        \toprule
        Number\# & $k_1$ & $k_2$ & $k_3$ & $k_4$ \\
        \midrule
        1 (ground truth) & $1.000$ & $1.000$ & $1.000$ & $1.000$ \\
        2                & $1.722$ & $0.636$ & $1.301$ & $0.701$ \\
        3                & $1.999$ & $1.000$ & $0.500$ & $1.000$ \\
        4                & $2.640$ & $0.647$ & $0.813$ & $0.720$ \\
        \bottomrule
    \end{tabular*}
\end{table}

The quantities of interest herein are the frequency response function (FRF) of roof-top acceleration with respect to base excitation.
We define a simulator as $h : \bm{\theta} \mapsto \mathbf{x}$, where $\mathbf{x} \in \mathbb{R}^{1024}$ represents the logarithm of the absolute FRF, $\log |H(f)|$, evaluated at the frequency points $f_j = \{0.00, 0.02, ..., 20.46 \}$ Hz with additive observation noise.
This reads
\begin{align}
    x_j = \log |H(f_j)| + \epsilon_j, \quad \epsilon_j \sim \mathcal{N}(\cdot \mid 0, \sigma_\epsilon^2),
\end{align}
for $j = 1, ..., 1024$.
We set $\sigma_\epsilon = 0.2$.

\begin{figure}[!t]
    \centering
    \includegraphics[width=1.00\linewidth]{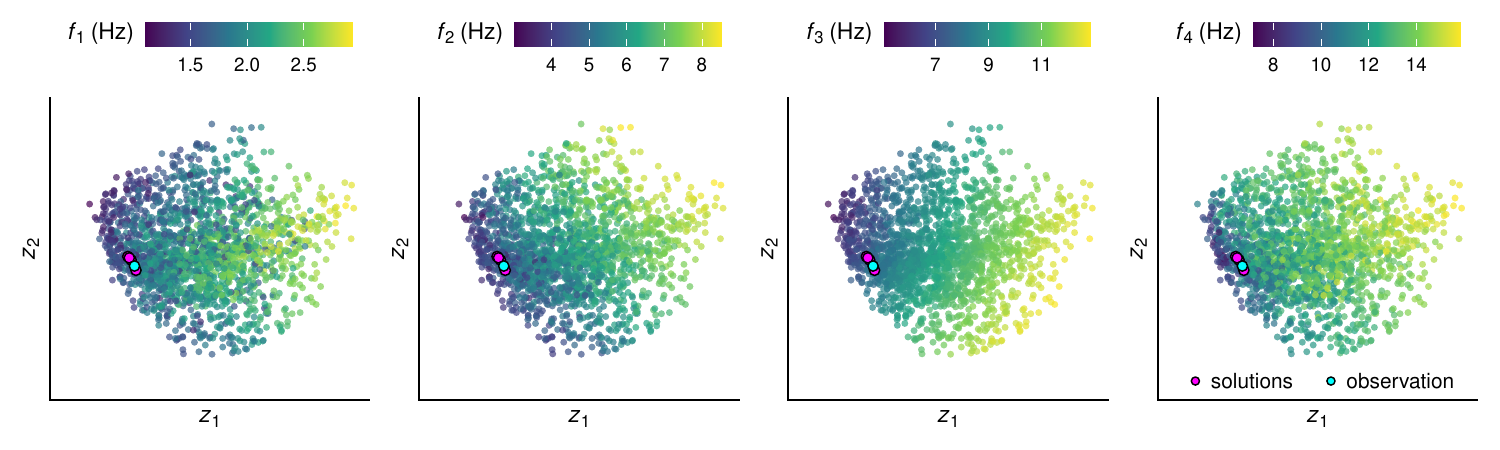}
    \caption{Visualization of the latent space of the trained MVAE model in the benchmarking example.}
    \label{fig:latent}
\end{figure}

\subsection{Training of MVAE}\label{sec:num-training}

We draw $10^5$ samples of $\bm{\theta}^{(n)}$ from the prior and run the simulator to acquire responses $\mathbf{x}^{(n)} = h(\bm{\theta}^{(n)})$ as well as contaminated ones $\tilde{\mathbf{x}}^{(n)} = \mathbf{x}^{(n)} + \bm{\epsilon}^{(n)}$, yielding the training dataset $\{(\bm{\theta}^{(n)}, \mathbf{x}^{(n)}, \tilde{\mathbf{x}}^{(n)})\}_{n=1}^{10^5}$.
This is randomly split into training and validation sets, comprising 90\% and 10\% of the samples, respectively.
Based on these, we train an MVAE architecture consisting of $(\phi_\mathbf{x}, \phi_{\bm{\theta}}, \eta)$.
All neural networks include fully connected (FC) layers and residual blocks composed of convolutional layers.
The dimension of the latent space is set to $n_z = 8$.
For further details on the architecture, we refer the readers to our GitHub repository, \url{https://github.com/taroyaoyama/lsbi-smc/}.
To optimize these networks, we adopt the Adam algorithm with a learning rate of $10^{-3}$ and batch size of 256.
The number of training epochs is determined using the early stopping strategy, whereby the training is stopped when the validation loss does not improve for 20 epochs.
The entire training procedure is implemented using a Python package, PyTorch.

We investigate the geometric structure of the latent space learned by the trained MVAE.
For illustration, we draw prior samples $\bm{\theta}^{(n)} \sim p(\bm{\theta})$ and map them to the latent space through the encoder $q_{\phi_{\bm{\theta}}}(\mathbf{z} \mid \bm{\theta})$.
We then apply principal component analysis (PCA) to the mean latent vectors $\bm{\mu}_{\phi_{\bm{\theta}}} \in \mathbb{R}^8$ for visualization.
The resulting two-dimensional projections are shown in Figure~\ref{fig:latent}.
In each panel, the color indicates the corresponding natural frequency $f_m ~ (m = 1, ..., 4)$.
The projections are organized according to the natural frequencies, implying that the learned latent space successfully captures the intrinsic features of the FRFs.
We additionally annotate the four equivalent solutions listed in Table~\ref{tab:sols}, mapped into the latent space via the encoder $q_{\phi_{\bm{\theta}}}(\mathbf{z} \mid \bm{\theta})$, as well as the observation generated from the ground truth $\bm{\theta} = \{1,1,1,1\}^\top$, which is mapped through the encoder $q_{\phi_{\mathbf{x}}}(\mathbf{z} \mid \mathbf{x})$.
Although these equivalent solutions are widely separated in the original parameter space, they all cluster near the observations in the latent space.
This indicates that the MVAE successfully retrieves the shared latent representation between parameters $\bm{\theta}$ and responses $\mathbf{x}$ even under the limited parameter identifiability.

\subsection{Results of Bayesian inference}

Using the MVAE-based likelihood approximation $\widehat{L}_{\phi_\mathbf{x},\phi_{\bm{\theta}}}(\bm{\theta}; \mathbf{x})$, we draw the posterior samples from $\widehat{p}(\bm{\theta} \mid \mathbf{x}_\mathrm{obs}) \propto \widehat{L}_{\phi_\mathbf{x},\phi_{\bm{\theta}}}(\bm{\theta}; \mathbf{x}_\mathrm{obs}) \, p(\bm{\theta})$.
We benchmark two sampling algorithms: SMC and No-U-Turn Sampler (NUTS) \citep{hoffmanMD2014}.
NUTS is the state-of-the-art MCMC algorithm and a variant of Hamiltonian Monte Carlo (HMC) \citep{duaneS1987,nealRM2011}.
HMC \citep{duaneS1987, nealRM2011} leverages Hamiltonian dynamics to construct gradient-informed proposals using a leapfrog integrator, leading to high acceptance rates and efficient exploration.
NUTS adaptively determines the trajectory length, eliminating the need for manual tuning of integration parameters.
As these methods require gradients of the log-likelihood, we compute them efficiently using automatic differentiation.

In this benchmark test, we run a single NUTS chain for 60,000 iterations, discarding the first 10,000 iterations as burn-in.
Note that the number of MCMC iterations does not coincide with the number of likelihood evaluations, since each iteration includes multiple leapfrog steps, each of which requires evaluations of the likelihood function and its gradient.
The implementation is based on a Python package, Pyro.
For SMC, we consider three population sizes $N_\mathrm{s} \in \{1000, 2000, 4000\}$, while fixing the number of MCMC iterations per round to 10.
These configurations are chosen such that total number of likelihood evaluations (denoted by $N_\mathrm{eval}$) of $\widehat{L}_{\phi_\mathbf{x},\phi_{\bm{\theta}}}(\bm{\theta}; \mathbf{x})$ is comparable between the SMC and NUTS algorithms.

To quantify the performance of the sampling algorithms, we employ the maximum mean discrepancy (MMD) \citep{grettonA2012}, a kernel-based metric that measures the discrepancy between two sample distributions.
As a reference, we generate posterior samples using SMC with a large population size of $N_\mathrm{s} = 5 \times 10^4$.
For each inference setting, we compute the MMD between the resulting posterior samples and this reference set; that is, a lower MMD value indicates higher inference performance.
In the MMD calculation, we only use 5000 samples randomly selected from the reference set for computational efficiency.

\begin{figure}[!t]
    \centering
    \includegraphics[width=1.00\linewidth]{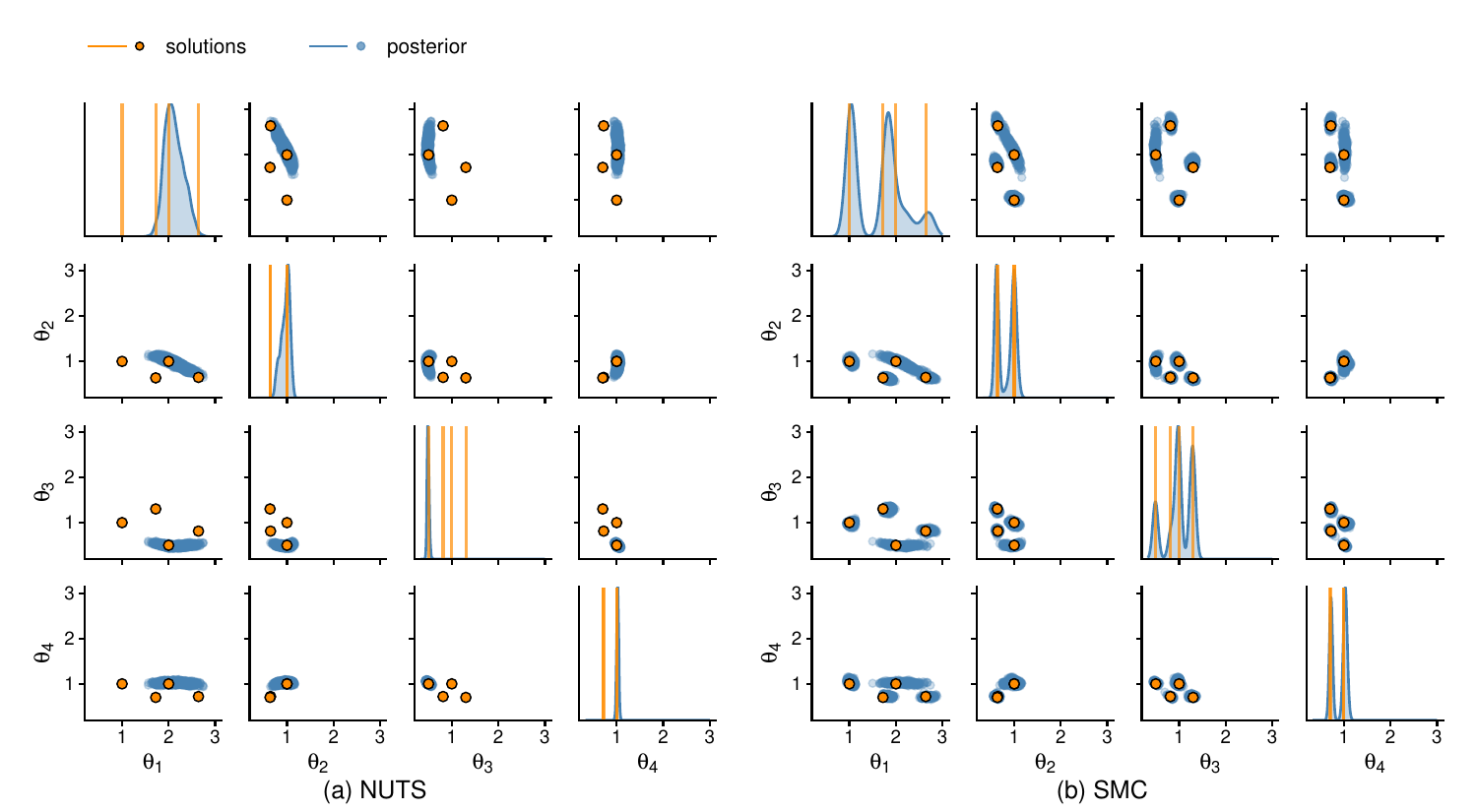}
    \caption{The distribution of posterior samples in the benchmarking example ($N_\mathrm{s} = 2000$).}
    \label{fig:post_4dof}
\end{figure}

\begin{figure}[!t]
    \centering
    \includegraphics[width=1.00\linewidth]{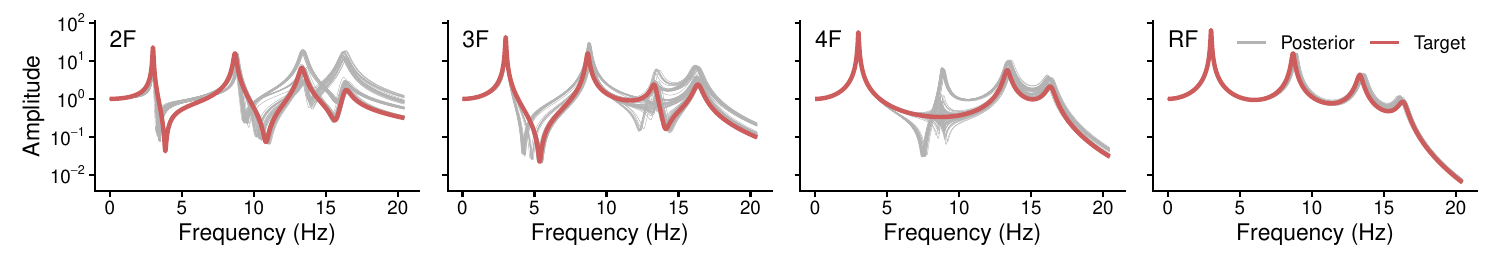}
    \caption{Posterior predictions of FRF in the benchmarking example ($N_\mathrm{s} = 2000$).}
    \label{fig:pred_4dof}
\end{figure}

\begin{table}[t]
    % from work_260222c.ipynb
    \centering
    \caption{Results in the 4DOF benchmark example for each sampling algorithm (summary of independent 10 runs).}\label{tab:results}
    \fontsize{8truept}{10truept}\selectfont
    \begin{tabular*}{\textwidth}{@{\extracolsep{\fill}}lrrrrr}
        \toprule
        ~ & ~ & ~ & \multicolumn{2}{c}{MMD} & ~ \\ \cmidrule{4-5}
        Algorithm & $N_\mathrm{s}$ & $N_\mathrm{eval}$ & mean    & std.    & Wall-clock time (sec) $^\dagger$ \\ \midrule
        SMC       & $ 500$         &  $91000$          & $0.086$ & $0.038$ &  $0.8$                           \\
                  & $1000$         & $182000$          & $0.077$ & $0.034$ &  $0.8$                           \\
                  & $2000$         & $362000$          & $0.051$ & $0.019$ &  $0.8$                           \\
                  & $4000$         & $724000$          & $0.040$ & $0.015$ &  $0.8$                           \\
        NUTS      & $-$            & $558871$          & $0.629$ & $0.159$ & $1782$                           \\
        \bottomrule
    \end{tabular*}
    $^\dagger$~These do not include the elapsed time required for dataset construction and MVAE training. \hfill~
\end{table}

Table~\ref{tab:results} summarizes the performance for different inference settings averaged over independent ten runs, while the likelihood estimator $\widehat{L}_{\phi_\mathbf{x},\phi_{\bm{\theta}}}(\bm{\theta}; \mathbf{x})$ is common across all runs and settings.
SMC, for all considered values of $N_\mathrm{s}$, exhibits superior performance compared to NUTS in terms of MMD.
Notably, SMC also requires substantially less wall-clock time than NUTS, highlighting the computational efficiency enabled by GPU-parallelized implementation.

Figure~\ref{fig:post_4dof} presents a scatter plot matrix of posterior samples obtained from NUTS and SMC ($N_\mathrm{s} = 2000$).
The marginal posterior distributions are estimated via kernel density estimation (KDE) and displayed along the diagonal.
While NUTS fails to fully explore the four posterior modes associated with the equivalent solutions, SMC successfully captures all modes.
Figure~\ref{fig:pred_4dof} presents the predictive distribution of FRFs on all floors, which are simulated from posterior samples in SMC ($N_\mathrm{s} = 2000$), as well as the target FRF for the ground-truth parameter $\bm{\theta} = \{1,1,1,1\}^\top$.
For the roof-top response (RF), the posterior predictive distributions closely align with the observed target, indicating that the inference successfully aligns with the measurement.
For the unobserved floors (2F, 3F, and 4F), the predicted FRFs exhibit pronounced multimodality while still covering the ground-truth responses.
These results demonstrate that the proposed LSBI-SMC approach maintains a robust exploration capability, even in severely ill-posed inference problems characterized by multimodal posteriors,
highlighting its suitability for model updating under practical sensing constraints.

\section{Application to RC building structures under seismic excitation}

\begin{figure}[!t]
    \centering
    \includegraphics[width=0.5\linewidth]{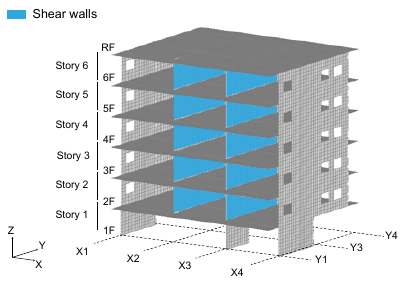}
    \caption{The finite element model of the target specimen in \cite{nied2015,sugimotoK2017}.}
    \label{fig:femodel}
\end{figure}

\subsection{Target structure and data}

The target structure is a scaled specimen of a six-story RC wall-frame building structure from the E-Defense shaking table test \citep{nied2015,sugimotoK2017}.
Figure~\ref{fig:femodel} displays the developed FE model for the target specimen.
The specimen has shear walls in the X2 and X3 frames and has non-structural walls in the X1 and X4 frames.
We specifically focus on the loading test termed Case \#1--2 in \cite{sugimotoK2017} under white noise excitation, and use acceleration measurements only in the X-direction.
In this study, response time histories from multiple sensors installed on each floor are averaged in the time domain and transformed into the logarithm of the absolute value of FRFs with respect to base acceleration.
Namely, we consider a $N_\mathrm{freq} \times N_\mathrm{story}$ data tensor for $\mathbf{x}_\mathrm{obs}$,
where $N_\mathrm{freq} = 1024$ is the number of considered frequency points and $N_\mathrm{story} = 6$ corresponds to all six floors (2F, 3F, ..., RF).

The developed FE model has shell elements for shear walls, and slabs and beam elements for beams and columns.
The fixed boundary conditions are introduced at the bottom surfaces of the columns and walls in the first story.
FE analyses are conducted using a commercial software, TDAP III \citep{tdap3}.
We parameterize the Young's modulus of concrete for the $i$th story as $E_i = \theta_i E_0$, where $E_0 = 2.44 \times 10^4$ (N/mm$^2$) denotes a nominal value, and $\theta_i$ represents a calibration coefficient.
This configuration results in the parameter vector to be estimated as $\bm{\theta} = \{\theta_1, ..., \theta_6\}^\top \in \mathbb{R}^6$.
Poisson ratio is set as 0.2.
The constant damping ratio $\zeta$ is assumed for all vibration modes.
The damping ratio $\zeta$ is considered uncertain within the range $[0.0066,0.06]$, and herein is not treated as an inference parameter, as it is assumed to be evaluated separately using system identification or related techniques.
Instead, when generating the training dataset, $\zeta$ is sampled from a uniform distribution over this range to reflect its uncertainty.

To investigate the effect of sensor placement on the parameter estimation uncertainty, we compare two observation scenarios.
Case 1 corresponds to the situation where the FRFs at all stories are obtained, whereas Case 2 assumes that only the roof-top FRF is available.
In Case 2, the parameter vector $\bm{\theta}$ is expected to be only partially identifiable due to the limited observations.

\begin{figure}[!t]
    \centering
    \includegraphics[width=0.5\linewidth]{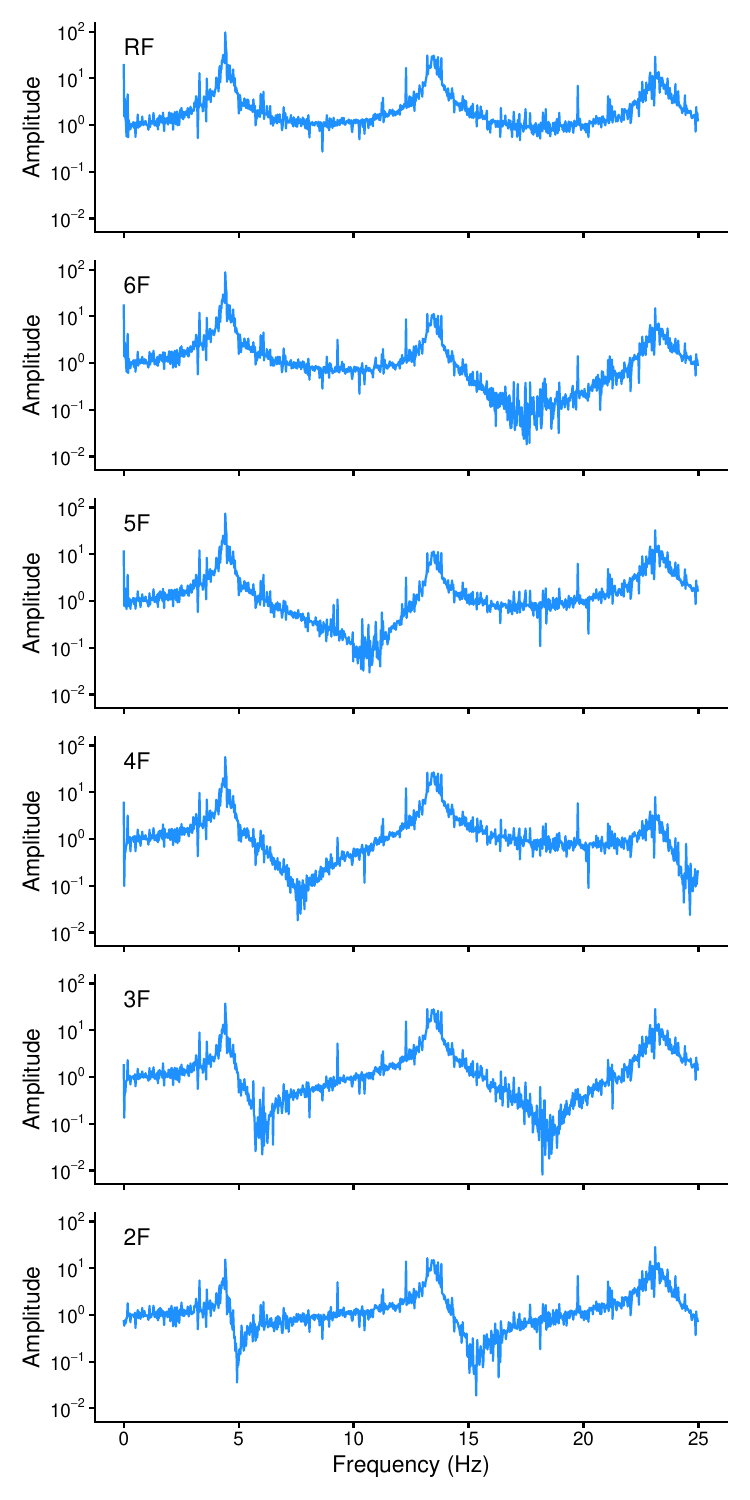}
    \caption{Observed frequency response functions in the experimental data \citep{nied2015,sugimotoK2017}.}
    \label{fig:exp-obs}
\end{figure}

\begin{figure}[!t]
    \centering
    \includegraphics[width=1.0\linewidth]{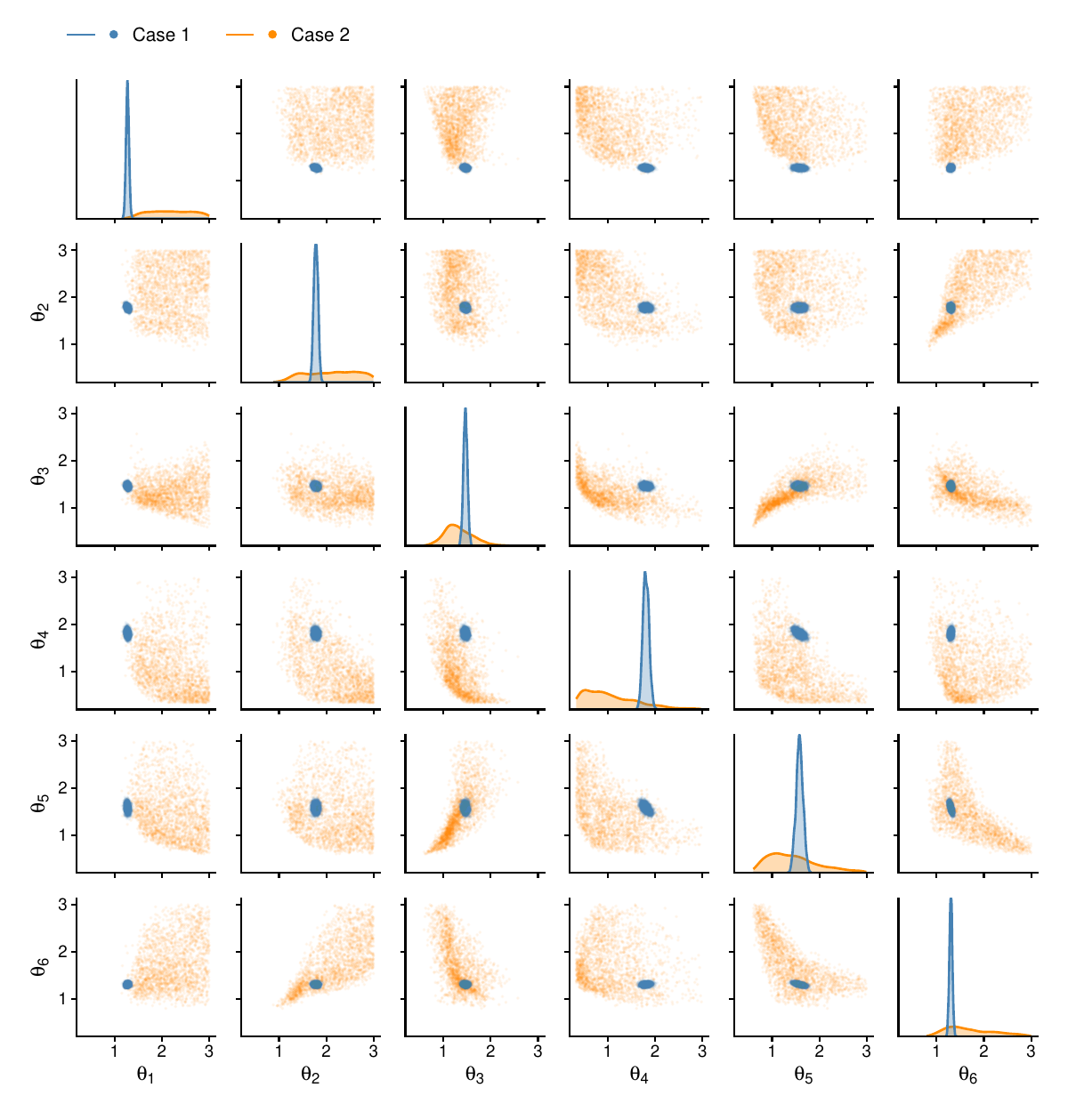}
    \caption{Inferred posterior samples in the experimental study for Case 1 (fully observed) and Case 2 (only the roof-top FRF observed).}
    \label{fig:exp-post}
\end{figure}

\begin{figure}[!t]
    \centering
    \includegraphics[width=1.0\linewidth]{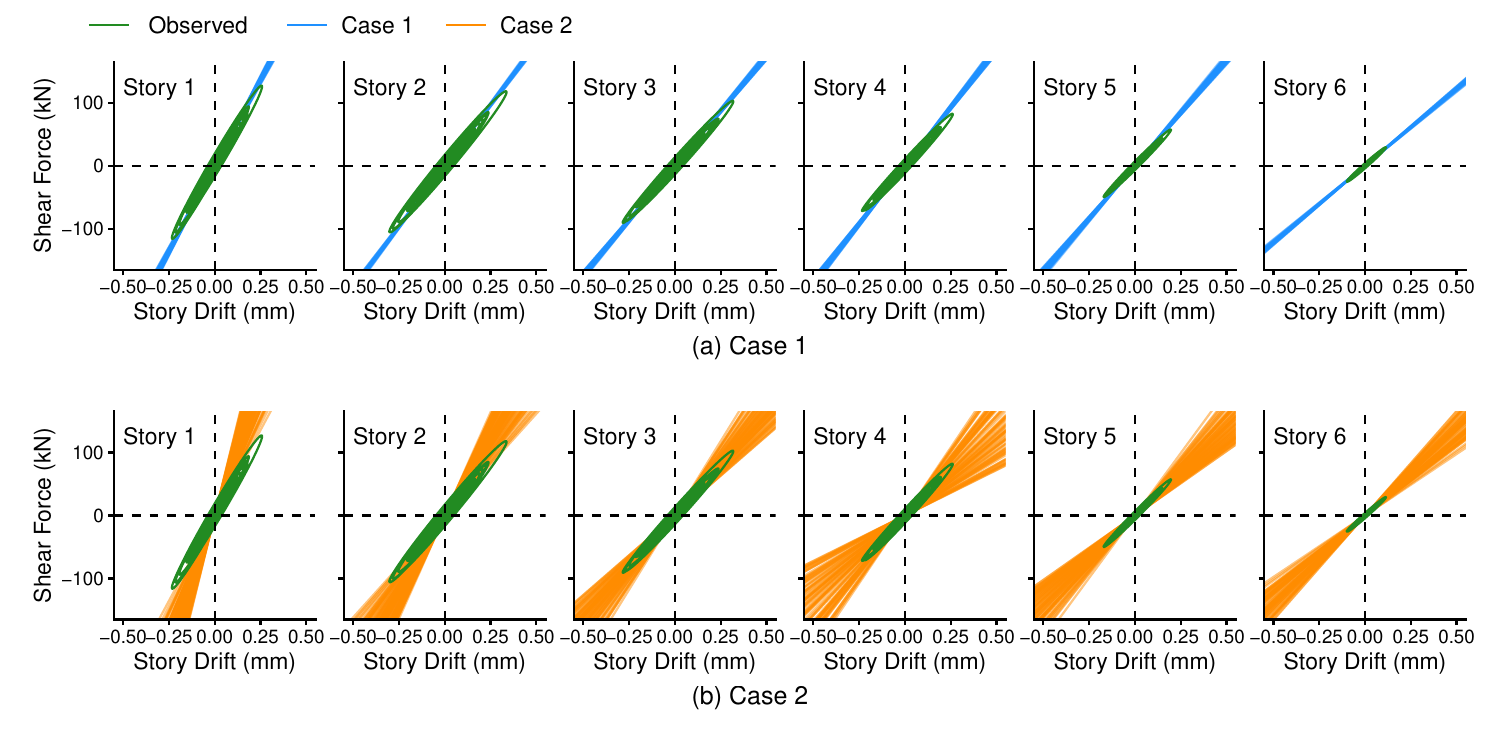}
    \caption{Predicted $Q$--$d$ relationships for the 10\%--scaled JMA Kobe excitation.}
    \label{fig:exp-stif}
\end{figure}

\begin{figure}[!t]
    \centering
    \includegraphics[width=1.0\linewidth]{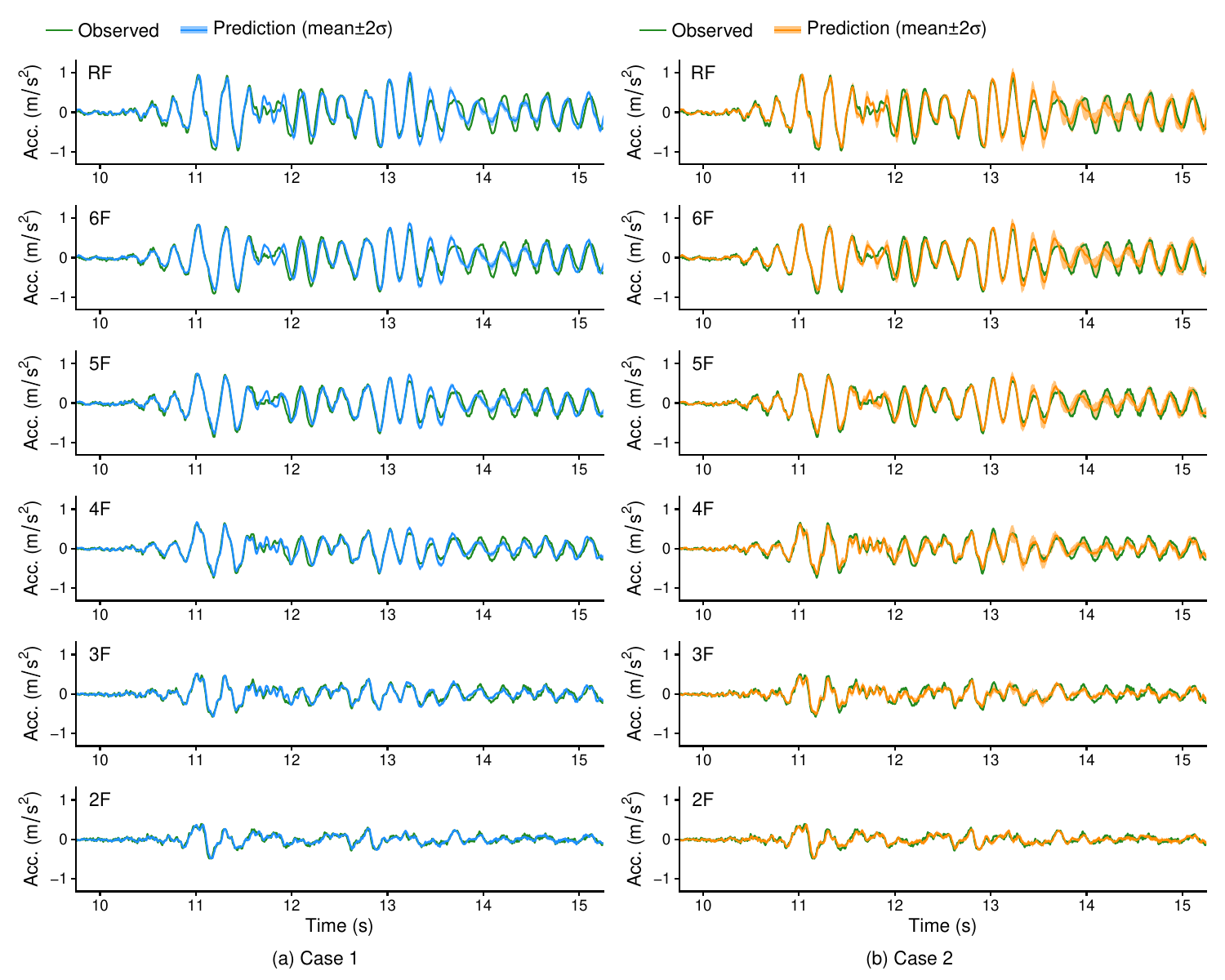}
    \caption{Posterior distributions of time history responses for the 10\%--scaled JMA Kobe excitation.}
    \label{fig:exp-pred}
\end{figure}

\subsection{Training of MVAE}

We adopt the same MVAE architecture as that described in \ref{sec:num-training}.
For the training dataset, we generate $N_\mathrm{train} = 10^5$ samples of $\bm{\theta}$ from the uniform prior distribution $\bm{\theta} \sim \mathcal{U}(\cdot \mid [1/3, 3]^6)$.
Linear dynamic FE analyses under white-noise excitation are conducted using the mode superposition method to construct the synthetic dataset $\{(\bm{\theta}^{(n)}, \mathbf{x}^{(n)}, \tilde{\mathbf{x}}^{(n)})\}$, where $\tilde{\mathbf{x}}^{(n)}$ is created by adding Gaussian noise to $\mathbf{x}^{(n)}$ according to $\mathcal{N}(\cdot \mid 0, 0.2^2)$.
Ninety percent of the dataset is allocated for training, with the remaining ten percent for validation.
For training, $\theta_i$ is rescaled to the range $[0, 1]$, while $\{\mathbf{x}_i\}$ is standardized to have zero mean and unit variance across the entire training dataset.
The MVAE is trained using the Adam optimizer with a batch size of 256 and learning rate of $5 \times 10^{-4}$.
Similarly to \ref{sec:num-training}, the training is stopped with no improvement in 20 epochs.

\subsection{Results of Bayesian inference}

For both Cases 1 and 2, we perform LSBI-SMC with a population size of $5 \times 10^4$, where the prior distribution is defined as $\bm{\theta} \sim \mathcal{U}(\cdot \mid [1/3, 3]^6)$.
Figure~\ref{fig:exp-post} presents a scatter plot matrix of 2,000 randomly selected posterior samples.
The marginal posterior distributions are estimated via KDE and shown along the diagonal.
The posterior distribution in Case 2 exhibits a broader spread, reflecting the limited identifiability induced by sparse measurements.
In contrast, the posterior distribution in Case 1 is contained within that of Case 2 but substantially narrower, suggesting that increasing the number of instrumented floors improves parameter identifiability and reduces the uncertainty in parameter estimation.

To validate the inferred posteriors, we predict seismic responses under a different excitation (namely the 10\%--scaled JMA Kobe record), corresponding to Case \#1--3 defined in \cite{sugimotoK2017}.
Figure~\ref{fig:exp-stif} presents straight lines (shown in blue and orange) whose slopes correspond to the predicted story stiffness obtained from static FE analyses based on 100 posterior samples.
These predictions are compared with the observed $Q$--$d$ relationships (shown in green), where the story shear force $Q$ (kN) is computed from acceleration responses and story masses, and the story drift $d$ (mm) is measured by laser displacement sensors.
In Case 1, the estimated story stiffness shows good agreement with the observed $Q$--$d$ relationships for all stories.
In Case 2, the predicted story stiffness exhibits a wider spread across all stories, reflecting increased uncertainty due to limited sensor placement.
For the first story, the predictions deviate from the observations, reflecting the discrepancy of the posterior distribution of $\theta_1$ between Case 1 and Case 2, as shown in Figure~\ref{fig:exp-post}.
Nevertheless, the predicted $Q$--$d$ relationships remain generally consistent with the observations for most stories, indicating that the proposed method captures the overall structural response even under reduced observability.
Figure~\ref{fig:exp-pred} presents the predictive distributions of acceleration responses obtained from dynamic FE analyses using 100 posterior samples.
For these simulations, the damping ratio is fixed at $\zeta = 0.02$, which is estimated independently based on system identification.
The $\pm 2\sigma$ intervals estimated from these samples are shown together with the observed responses.
For both cases, the predictions agree well with the observations, particularly around 11~s where large amplitudes are observed.
Although Case 2 exhibits greater predictive uncertainty than Case 1, the observed responses are still well captured, demonstrating the robustness of the proposed model updating framework under sparse sensor placement and limited parameter identifiability.

\section{Conclusions}

This study proposed a machine learning-aided Bayesian FE model updating framework for RC building structures under seismic loading, utilizing a latent space learned by a multimodal variational autoencoder (MVAE).
The proposed approach enables efficient likelihood approximation in a low-dimensional space and allows amortized inference, thereby eliminating the need for additional FE simulations during posterior sampling.
Furthermore, by incorporating a sequential Monte Carlo (SMC) sampler with GPU-based parallelization, the framework achieves high computational efficiency while maintaining robustness against multimodal and complex posterior distributions.

The proposed framework, LSBI-SMC, was validated through both numerical benchmarking and experimental data from shake-table loading tests on RC building structures.
In numerical benchmarking, we considered a shear-type 4DOF building structure with acceleration measurements only at ground and roof-top levels, thus exhibiting multimodal posterior distributions in stiffness parameters due to multiple equivalent solutions compatible with identical observations. 
The results demonstrated that the proposed SMC-based approach successfully captures all posterior modes, while achieving significantly higher computational efficiency compared with a state-of-the-art MCMC method, the No-U-Turn Sampler (NUTS).
In the experimental study, we applied the proposed framework to a high-fidelity FE model of an RC building structure, constructed using shell and beam elements.
We compared two observation scenarios: one in which only the roof-level frequency response function (FRF) is available, and another in which responses from all floors are observed.
The inferred posterior distributions reasonably reflected the degree of sensor availability.
The predictive performance of the updated FE model was further evaluated using a different seismic excitation, showing that the predicted story stiffness and time-history responses were in good agreement with the observations, thereby validating the proposed framework.

In future work, we will examine the applicability of the proposed framework to high-dimensional parameter estimation problems.
We will also investigate extensions to cases where the form of observation or modeling errors is not known a priori.

\section*{Acknowledgements}

This study was conducted as part of the research project in partnership between the University of Tokyo and Taisei Corporation.

\section*{Declaration of generative AI and AI-assisted technologies in the manuscript preparation process}

During the preparation of this work the authors used ChatGPT in order to improve the readability and language of the manuscript.
After using this tool, the authors reviewed and edited the content as needed and take full responsibility for the content of the published article.

\bibliographystyle{elsarticle-harv}
\bibliography{cas-refs.bib}

\end{document}